# Fusion Learning from Dynamic Functional Connectivity: Combining the Amplitude and Phase of fMRI Signals to Identify Brain Disorders

Jinlong Hu *, Jiatong Huang, Zijian Cai

Guangdong Provincial Key Laboratory of Multimodal Big Data Intelligent Analysis, School of Computer Science and Engineering, South China University of Technology, Guangzhou, China

jlhu@scut.edu.cn

## ABSTRACT

Dynamic functional connectivity (dFC) derived from resting-state functional magnetic resonance imaging (fMRI) has been extensively utilized in brain science research. The sliding window correlation (SWC) method is a widely used approach for constructing dFC by computing correlation coefficients between amplitude time series of signals from pairs of brain regions. In this study, we propose an integrated approach that incorporates both amplitude and phase information of fMRI signals to improve the detection of brain disorders. Specifically, we introduce a multi-scale fusion learning framework, namely MSFL, which leverages two complementary dFC features derived from SWC and phase synchronization (PS). Here, SWC captures amplitude correlations, while PS measures phase coherence within dFC. We evaluated the efficacy of MSFL in classifying autism spectrum disorder and major depressive disorder using two publicly available datasets: ABIDE I and REST-meta-MDD, respectively. The results indicate that MSFL significantly outperforms existing comparative models. Moreover, we performed model explanation analysis using the SHAP framework, which showed that both types of dFC features from SWC and PS contribute to detecting brain disorders.

## 1. Introduction

Functional magnetic resonance imaging (fMRI) is a non-invasive technique that provides valuable insights into the functional architecture of the brain by measuring spontaneous brain activity. Functional connectivity (FC) analysis can reveal how these areas of heightened activity work together, showing how the brain's functional networks interact with its structural connectivity patterns. Traditional static FC (sFC) assumes that connectivity is constant over time, thereby overlooking the dynamic nature of the brain. To address this limitation, dynamic functional connectivity (dFC) has been introduced to capture time-dependent fluctuations in brain network interactions, adding a critical temporal dimension to conventional FC analysis [1]. FC has been validated as a valuable analytical tool for investigating a wide range of brain disorders [2, 3].

In recent years, deep learning models have been successfully applied to analyze FC data in identifying brain disorders, learning from sFC data such as GAT-LI [4] and BrainNPT [5], and from dFC data such as BrainNetFormer [6], MT-STN [7], and TMN [8]. Most current research employs sliding window correlation (SWC) to derive dFC features, primarily focusing on the correlation of signal amplitudes. Recently, SWC has become one of the most commonly combined approaches with deep learning techniques to investigate brain dynamics. The core principle of SWC involves capturing transient fluctuations in brain functional connectivity by applying a fixed-size sliding window, enabling a more detailed understanding of the temporal evolution of brain network interactions. While features generated through SWC provide valuable insights into neural synchronization, they have certain limitations [9, 10]. Among the various methodologies used to derive dFC, PS stands out as a particularly important technique. PS measures the phase coherence between neural signals from different brain regions. This method can detect aspects of functional coupling that SWC might overlook, making SWC and PS complementary analytical approaches. Conceptually, combining these two distinct feature types has the potential to provide a more comprehensive and robust characterization of brain dynamics, thereby improving the accuracy and depth of subsequent analyses.

In this study, we introduce a multi-scale fusion learning framework, termed MSFL, which integrates features derived from SWC and PS methods. The MSFL framework employs a novel feature fusion architecture based on a cross-difference attention (CDA) mechanism, explicitly capturing both the similarities and, importantly, the distinctions between SWC and PS features. Simultaneously, a multi-scale convolutional network was designed to effectively extract dFC patterns across various temporal scales. We evaluated MSFL on two large-scale public datasets demonstrating that MSFL outperforms current state-of-the-art approaches. Furthermore, a model explanation analysis substantiates the complementary benefits of combining SWC and PS features, as well as identifies the key functional connectivity features that contribute most to the classification models' decisions.

## 2. Related works

FC features derived from resting-state fMRI data offer a fundamental method for quantifying statistical relationships between distinct brain regions. These features have been demonstrated to improve the performance of deep learning models and have achieved significant success across various classification tasks. Building on this framework, dFC features incorporate a temporal dimension, enabling effective characterization and analysis of fluctuations in brain connectivity over time.

During the past few years, deep learning models have also been successfully applied to the analysis of dFC data. For example, Fan et al. [11] treated dFC as time-series data and developed an end-to-end deep learning model that combines a Convolutional Neural Network (CNN) with a Long Short-Term Memory (LSTM) network for tasks such as gender classification and intelligence prediction. Wang et al. [12] proposed an attention-based Graph Convolutional Network (GCN) to extract spatiotemporal features from dFC, capturing key brain network patterns associated with specific cognitive states or diseases. Deng et al. [13] used the Transformer architecture based on ST-Transformer to extract spatiotemporal features from dFC. MT-STN [7] is an effective spatiotemporal learning model for dFC data that identifies major depressive disorder (MDD) from healthy controls (HC) using a stacked neural network architecture combining a multi-layer perceptron (MLP) and Transformer modules. BrainPST [14] integrates spatial and temporal information by stacking two Transformers: one for learning snapshot networks and another for learning sequences of functional connections, contructing a fundational model with pretraining on unlabeled dFC data.

Although the aforementioned studies have demonstrated the superiority of dFC in brain network classification, these dFC-based methods primarily rely on the SWC technique to extract dFC features. SWC focuses on the correlation between brain regions but presents several challenges. Hillman et al. [15] and Maltbie et al. [16] have indicated that the performance of the SWC method is highly sensitive to parameter choices such as window length, window offset, and window type. Moreover, SWC provides a poor estimation of the true correlation for each brain state and lacks a strong association with the actual connectivity, especially during abrupt state transitions. To overcome the limitations of relying on a single feature type, some studies have proposed dual-channel models that utilize feature fusion to enhance classification performance. For example, Huang et al. [17] fused SWC data from multiple brain templates; Chen et al. [18] combined SWC with dynamic effective connectivity features; and Kan et al. [19] integrated SWC with static functional connectivity features. Despite these varied approaches, developing effective classification models from complex spatiotemporal dFC data remains a significant challenge.

PS, another method for dFC assessment, primarily focuses on the synchronicity between brain regions by measuring whether the time series of different regions exhibit phase coherence. Pedersen et al. [20] directly and systematically compared instantaneous phase synchrony analysis (IPSA)—a PS method—with SWC, finding a nonlinear relationship between them at the topological level. Specifically, when SWC showed a strong negative correlation (i.e., signals from two brain regions changing in opposite directions), IPSA could still indicate high phase synchrony, suggesting a form of functional synergy. Honari et al. [21] explored the methodology of PS to optimize its application in fMRI data. Their study highlighted the theoretical advantage of PS over sliding-window methods in temporal resolution, as PS provides an instantaneous assessment of functional connectivity. They also proposed new techniques to more accurately capture changes in phase relationships, demonstrating that PS can reveal complex neural dynamics that SWC might obscure or oversimplify, especially during transitions between positive and negative correlations. Conversely, while not directly comparing with PS, Xie et al. [22] provided important context for the applicability of SWC. They found that SWC with longer windows was highly effective in distinguishing between different cognitive task states. This suggests that SWC focuses on capturing longer-lasting functional connectivity patterns associated with specific cognitive states, contrasting with PS's emphasis on instantaneous, rapidly changing synchrony. This distinction underscores their complementary roles in capturing neural activity across different time scales. Finally, Gandhi et al. [23] provided direct evidence of this difference and complementarity. They compared the two methods in identifying individualized brain fingerprints and found that SWC performed better on resting-state data, whereas PS was more advantageous for task-state data (listening to music). This result strongly demonstrates that PS and SWC are not simply a matter of one being superior to the other; rather, they capture different aspects of the brain's functional coordination in different contexts. SWC may be better suited for capturing stable, intrinsic connectivity properties, while PS is more sensitive to dynamic, stimulus-driven synchronization events.

In summary, the two forms of dFC information derived from SWC and PS are highly relevant for characterizing dynamic connectivity and provide complementary insights into the brain's dynamic functional coordination. Conceptually, integrating these SWC- and PS-based measures may enhance the identification of brain disorders. To the best of our knowledge, this study is the first work to utilize both amplitude and phase information obtained from SWC- and PS-based dFC analyses for brain disorder identification.

# 3. Methods

## 3.1 Data Preprocessing

This study use dFC features from fMRI Blood Oxygen Level Dependent (BOLD) signals by SWC and PS techniques. The process of SWC and PS methods are shown as Figure 1.

(1) Sliding window correlation

The sliding window correlation (SWC) procedure involves applying a fixed-length temporal window to the fMRI time series data and calculating the correlation coefficient between two signals of interest within this window. The window is then shifted by a predetermined step size, and the correlation calculation is repeated iteratively until the entire time series is covered. For each windowed segment, the functional connectivity between pairs of time series from different brain regions is quantified. In the present study, Pearson correlation is used for feature extraction through SWC. The Pearson correlation coefficient, calculated between two time series from different brain regions, ranges from -1 to 1.

(2) Phase synchronization

Phase synchronization (PS) is an advanced technique used to assess synchrony between time series derived from distinct brain regions. The process involves extracting phase information from the BOLD signal using the Hilbert transform. First, the BOLD signal is preprocessed to remove high-frequency noise. Next, the Hilbert transform is applied to generate the analytic signal, enabling the computation of the instantaneous phase of each brain region's BOLD signal. To evaluate synchrony between two brain regions, the phase difference between their respective time series is calculated, and the Phase Locking Value (PLV) is employed as a metric to quantify the temporal consistency of these phase differences. The methodology for deriving phase synchronization features is illustrated in Figure 1b, where the PLV approach is used to analyze phase relationships across different brain regions and quantify their synchrony.

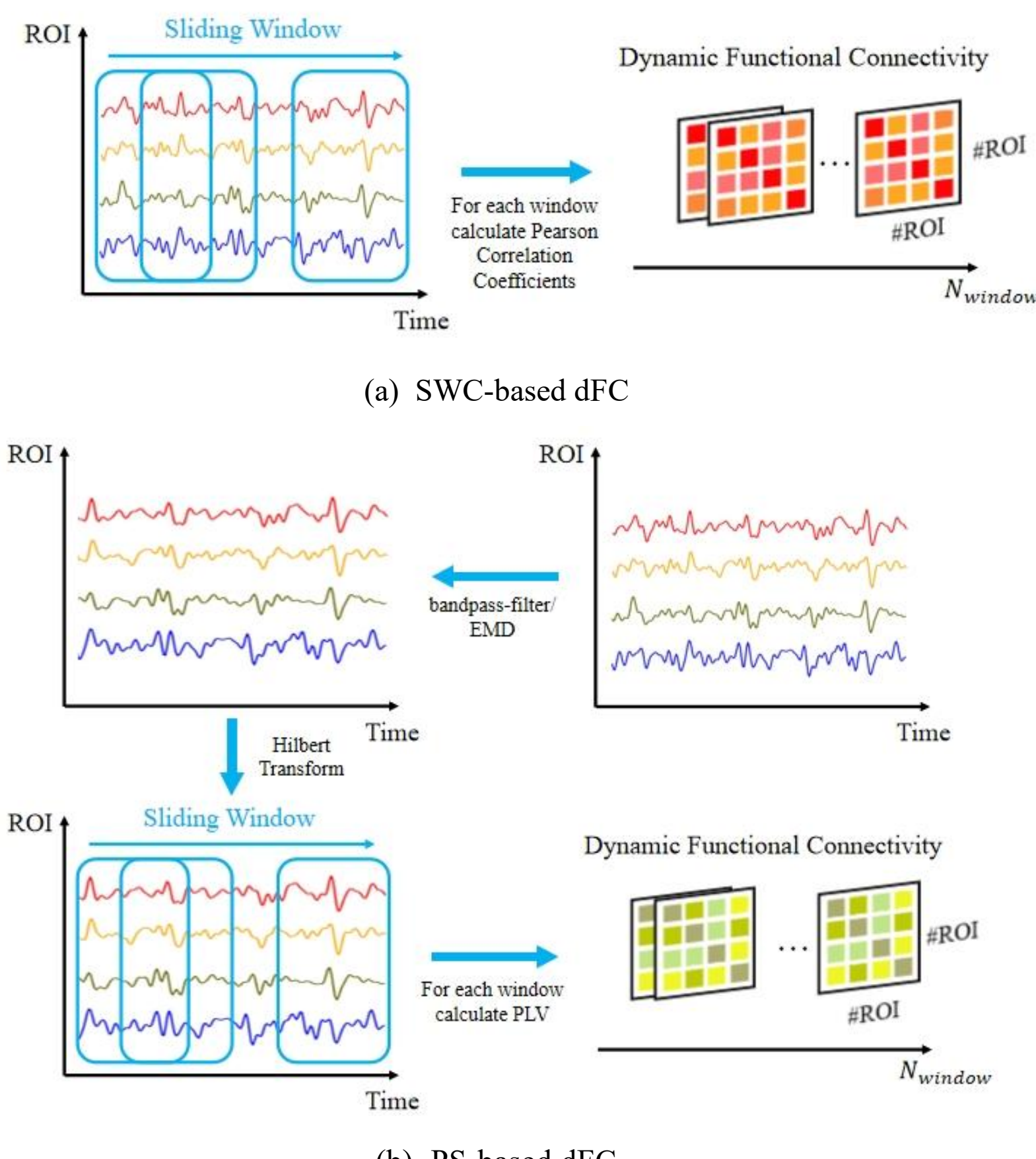

(a) SWC-based dFC

(b) PS-based dFC

Fig. 1 The extraction of dFC features from SWC and PS. (a) SWC-based dFC. (b) PS-based dFC.

### 3.2 The MSFL Framework

We introduce a multi-scale fusing learning framework, designated as MSFL, which combines SWC and PS features. The proposed MSFL framework utilizes a cross-attention mechanism to fuse these two distinct feature branches, thereby enhancing the representation of temporal and spatial information. The overall architecture of the MSFL model is illustrated in Figure 2 and consists of three main components: a Feature Fusion Module, a Multi-scale Convolutional Layer, and a classification head.

The model initially extracts SWC-based dFC and PS-based dFC features from resting-state fMRI data using their respective extraction techniques. These features are then subjected to dimensionality reduction and flattening before being input into the model's feature fusion module. The Feature Fusion Module includes two cross-attention components along with a Cross-Difference Attention (CDA) submodule, designed to enhance the interaction and integration of heterogeneous feature sets. The CDA submodule effectively identifies differences between two distinct feature types while simultaneously learning their inherent correlations through a cross-attention mechanism.

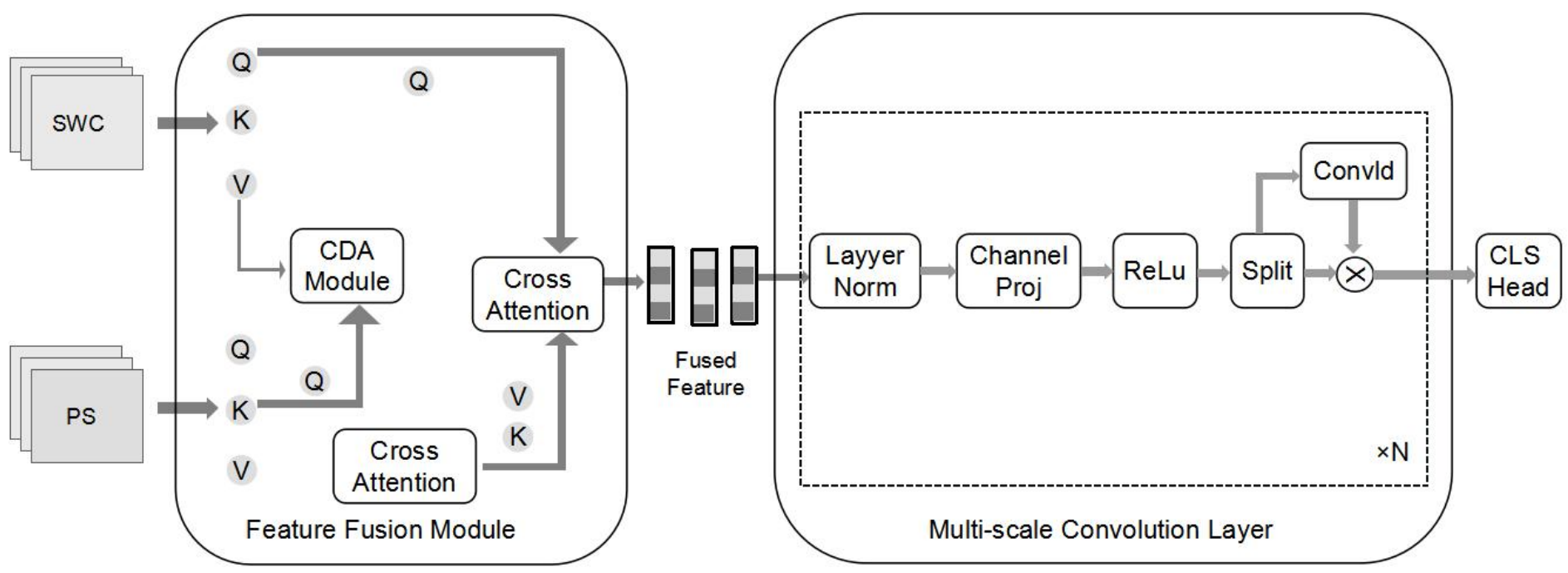

Fig. 2 The MSFL Framework

Following fusion, the integrated features are refined using a multi-scale convolution module that captures patterns across various temporal scales. Finally, the refined features are fed into a classification head, which produces the final classification results.

## 3.2 Model Input

For each subject, a time series $x \in R^{T \times NROI}$ is obtained, where *T* is the length of the time series and *NROI* is the number of Regions of Interest (ROIs). dFC is constructed using SWC and PS, respectively, yielding x_SWC, x_PS ∈ R^(N_windows × NROI × NROI). Since the FC matrix for each time window in the constructed dFC features is a high-dimensional symmetric matrix, its upper triangular part is first extracted and flattened to reduce redundant information. This results in x_SWC, x_PS ∈ R^(N_windows × D), where *D* is the feature dimension. A further dimensionality reduction is then applied to improve computational efficiency and the effectiveness of the feature representation.

## 3.3 SWC and PS Feature Fusion

As shown in Figure 3, the Cross-Difference Attention (CDA) module is designed to extract rich differential information between the PS and SWC features. this module employs the scaled dot-product attention mechanism. Three learnable weight matrices, WQ,WK, and WV, are used to transform the inputs into the Query (Q), Key (K), and Value (V) components of the attention mechanism, respectively. The attention weights are calculated from Q, K, and V using the following formula:

$$AttWeights(Q,K) = soft\max(\frac{QK^T}{\sqrt{d}}) \quad (1)$$

Where *d* represents the feature dimension of Q and K. The Softmax operation normalizes the dot product results to generate attention weights, which measure the similarity between different feature branches across various time windows. Higher attention weights indicate a stronger similarity between the corresponding windows.

After the attention weights are calculated, they perform a weighted sum on V to obtain the output of the attention module:

$$AttOutput(Q,K,V) = AttWeights(Q,K) * V \quad (2)$$

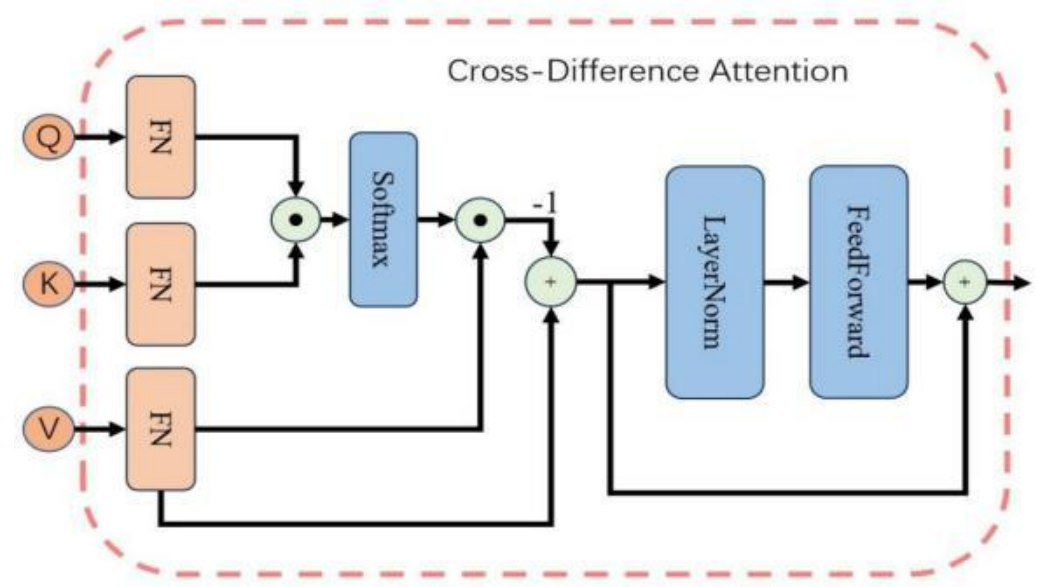

Fig. 3 The Architecture of the Cross-Difference Attention Module

In the CDA module, an additional step is performed where the attention output is subtracted from 1, thereby highlighting the differences between the features. This operation directs the model to focus more on the temporal variations and inconsistencies between the PS and SWC features. The calculation is as follows:

$$DiffOutput = 1 - AttOutput(Q, K, V) \tag{3}$$

In this step, the parts that originally had low attention scores—that is, the parts with high dissimilarity between the two features—will now produce a higher output. This adjustment enables the model to better focus on the differences between the two types of features. Subsequently, this difference output is added to V through a residual connection and then processed by LayerNorm and a feed-forward neural network to enhance the model's stability and learning capability.

The feature fusion module comprises one CDA module and two cross-attention based CA modules. The CDA module projects the SWC and PS features into Q_ps, K_swc, and V_swc using learnable matrices. Following the calculation steps described above, the output of the CDA module, denoted as h_CDA, is obtained.

$$h_{CDA} = CDA(Q_{PS}, K_{SWC}, V_{SWC}) \qquad h_{CDA} \in R_{N_{win} \times D} \tag{4}$$

Next, the first CA module maps h_CDA and the PS features into Q_hCDA, K_ps, and V_ps through learnable matrices. The output is then obtained via the cross-attention mechanism:

$$h_{CA1} = CA(Q_{hCDA}, K_{PS}, V_{PS}) \qquad h_{CA1} \in R_{N_{win} \times D} \tag{5}$$

The second CA module then maps h_CA1 and the SWC features into Q_swc, K_hCA1, and V_hCA1 through learnable matrices. The final output of the feature fusion module is obtained through the cross-attention mechanism.

$$h_{CA2} = CA(Q_{SWC}, K_{hCA1}, V_{hCA1}) \qquad h_{CA2} \in R_{N_{win} \times D} \tag{6}$$

### 3.4 Multi-scale Convolutional Module

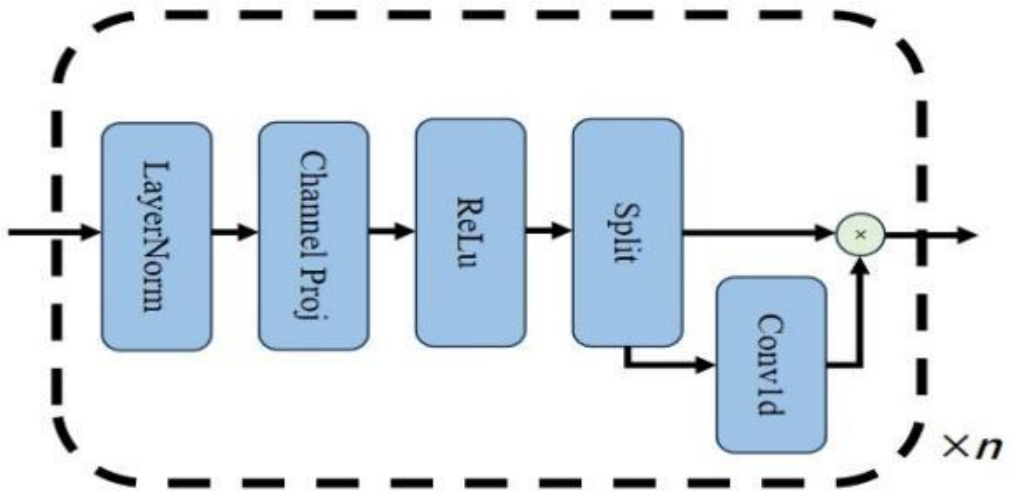

Fig. 4 Multi-scale Convolutional Module

As shown in Figure 4, after the SWC and PS feature fusion stage, the fused features are input into the multi-scale convolutional module for processing. The core design of this module involves progressively expanding the convolutional kernels to enlarge the receptive field, thereby capturing temporal dynamic features across multiple time scales. The multi-scale convolutional module employs a multi-layer structure, with each layer consisting of a uniform processing unit. In each successive layer, the kernel size of the one-dimensional convolution is gradually increased, enabling the model to capture both dynamic connectivity changes in dFC over short time scales and dynamic connectivity relationships over longer time scales. This strategy of gradual expansion ensures that the model can effectively process and integrate dFC features from different time scales, thereby enhancing its ability to understand the temporal dynamics of dFC.

At the beginning of each layer, the input features first pass through a channel projection layer (labeled *Channel Proj* in the Fig. 4), which doubles the channel dimension of the features. The expanded features are then evenly split into two parts by a Split layer: one part passes through a dilated one-dimensional convolutional layer to capture local, short-term change patterns in dynamic functional connectivity (dFC), while the other part remains unchanged to preserve the global information of dFC over longer time scales. After the convolution operation, the two parts are combined through element-wise multiplication, integrating local detailed information with global contextual understanding. As the dFC features propagate through the deeper layers of the multi-scale convolutional module, the model progressively learns more comprehensive temporal dFC features, effectively attending to brain functional connectivity dependencies across both long-term and various short-term time windows. This multi-scale approach enables the model to better capture the complex patterns in dFC data. The resulting features are then passed to the classification head.

### 3.5 Classification Head

The primary function of the classification head is to reduce the dimensionality of the high-dimensional features output by the multi-scale convolutional network and to produce the final classification result for the sample. This module consists of multiple fully connected layers that progressively reduce the feature dimensions, ultimately down to two. A Softmax function is then applied to normalize the output. The two dimensions correspond to the probability of the sample belonging to the patient group or the healthy control group, respectively.

## 4 Model Explanation

In the dFC classification model presented in this paper, the Shapley Additive Explanations (SHAP) method [24] is utilized to evaluate the impact of each dynamic functional connection on the model's prediction results. Specifically, the dFC matrix serves as the model input, and the SHAP method calculates an importance score for each functional connection in the classification task by determining the marginal contribution of each dFC feature under various conditions.

In the dFC classification task of this paper, the SHAP method is used to evaluate the impact of each dynamic functional connection on the model's prediction results. Specifically, the dFC matrix is used as the model input, and the SHAP method calculates the marginal contribution of each dFC feature under different conditions to derive an importance score for each functional connection in the classification task.

In the specific implementation process, SHAP is first applied to the trained SP-DSMAE classification model, and the feature contributions for each sample are calculated. For each dFC feature $x_i$, its SHAP value $\phi_i$ reflects its positive or negative contribution to the classification decision (i.e., patient group or healthy control group). By averaging the SHAP values across all samples, a global explanation can be obtained, which identifies the functional connections most critical to the classification task. Subsequently, the SHAP values are ranked to screen for high-contribution dFCs to further analyze their role in disease prediction. For example, if certain dFCs consistently have high SHAP contribution values in the autism patient group but low values in the healthy control group, these functional connections can be considered potentially disease-related.

Through this method, SHAP not only enhances the interpretability of the model's prediction results but also identifies functional connections that are crucial for the classification task, thereby deepening the understanding of disease-related brain functional characteristics.

## 5 Experiments and Result Analysis

### 5.1 Datasets

In this section, we evaluated the model's performance using two datasets: the REST-meta-MDD dataset [25] and the ABIDE I dataset [26]. The REST-meta-MDD dataset is one of the largest functional brain imaging datasets currently available, focusing on Major Depressive Disorder (MDD). It integrates data from 25 research groups across 17 medical institutions and universities in China, comprising a total of 1,300 patients with depression and 1,128 healthy controls. The ABIDE I dataset is one of the largest publicly available brain imaging datasets related to autism. It includes data from 539 individuals with autism and 573 healthy controls collected from 17 medical sites.

The experiments in this section utilized the Automated Anatomical Labeling (AAL) brain atlas to parcellate brain regions. After preprocessing the resting-state fMRI (rs-fMRI) data, the resulting BOLD signal time series for the brain regions were subjected to the following exclusion criteria: (i) Samples in which the BOLD signal of any brain region was consistently zero or showed no fluctuation after preprocessing were excluded. Such cases typically result from errors during data acquisition or preprocessing. (ii) Since PS features require a consistent sampling frequency, this experiment selected only samples with a sampling interval of 2000 ms. (iii) To maintain a consistent sample length, this experiment selected a length of 130 for the ABIDE I dataset and 200 for the REST-meta-MDD dataset. Samples shorter than these lengths were discarded. For samples exceeding these lengths, only the last 130 or 200 time points were retained, respectively.

### 5.2 Comparison Models and Experimental Setup

The comparison models include both single-channel and dual-channel dFC methods.

(i) Single-channel dFC Methods

The single-channel dFC methods include four models: CNNLSTM [27], FBLSTM [28], STAGIN [29], and BrainNetFormer [6]. CNNLSTM and FBLSTM are based on the long short-term memory (LSTM) model; the former uses a CNN to extract spatial structures and an LSTM to capture temporal features, while the latter employs a bidirectional LSTM to enhance its ability to extract temporal information. STAGIN, which is based on graph convolutional networks, treats dFC features as a spatiotemporal graph and excels at extracting both spatial and temporal information. BrainNetFormer, based on a cross-attention and Transformer architecture, improves representational learning by introducing spatial and temporal cross-attention modules.

(ii) Dual-channel dFC Methods

The dual-channel dFC methods include four models: SDCNN [17], DART [30], FE-STGNN [18], and MAGCN [31]. Among these, SDCNN fuses sFC and dFC features. Its static path employs high-resolution convolutional filters (i.e., filters with a large number of channels) on a single sFC adjacency matrix to capture static FC patterns, while its dynamic path uses low-resolution convolutional filters on each dFC adjacency matrix to capture time-varying FC patterns. DART is another method that integrates sFC and dFC features, learning spatial and temporal information from both paths based on the BrainTransformer architecture. FE-STGNN, which combines dFC with dynamic effective connectivity (DEC), is currently the best-performing dual-channel method, as it fuses two types of temporally rich features. MAGCN fuses dFC features extracted from different brain templates, learning the differential information between templates to improve classification accuracy.

Furthermore, this experiment configured the single-channel dFC methods in two ways: first, their performance was evaluated directly using their original SWC input without modification; second, our proposed feature fusion module was added to the front end of these models, converting them into dual-channel dFC methods to further validate the effectiveness of the feature fusion.

For the dual-channel dFC methods mentioned above, their inputs were replaced with the SWC and PS features from our experiment to ensure a fair comparison.

The dFC extraction for all the aforementioned methods is based on a sliding-window strategy. In our experiments, the sliding window size was set to 30, with a step size of 10. For the MSFL model, a fully connected layer was employed to reduce the dimensionality to 256. The multi-head self-attention mechanism in the attention-based fusion utilized 8 heads. The multi-scale convolutional module consisted of 4 layers, with one-dimensional convolutional kernel sizes of 3, 5, 7, and 9.

In the experiments, the AdamW optimizer was used with hyperparameters set to $\beta_1 = 0.9$, $\beta_2 = 0.999$, and a weight decay of $2 \times 10^{-4}$. The batch size was set to 32. The learning rate followed the One Cycle [32] scheduling policy, where it linearly increased from $1\times10^{-6}$ to $1\times10^{-5}$ during the first 10% of epochs, and then gradually decayed to $1\times10^{-9}$.

All models were trained using the cross-entropy loss function with five-fold cross-validation, with each fold trained for 200 epochs. This experiment employs five-fold cross-validation to evaluate the model's classification performance. Several metrics are selected for quantitative analysis, including accuracy (ACC), specificity (SPEC), sensitivity (SEN), and area under the curve (AUC).

### 5.3 Classification Results

The comparative experimental results are presented in Tables 1 and 2 for the ABIDE I and the Rest-Meta-MDD Dataset, respectively. For the input types, SWC denotes sliding-window correlation features, PS denotes phase synchronization features, and DEC denotes dynamic effective connectivity features. The results indicate that the MSFL model achieves the best performance in terms of ACC, SEN, and AUC. Furthermore, the dual-channel variants, created by incorporating the feature fusion module, outperform the original single-channel models in several performance metrics, demonstrating that the phase synchronization feature plays a significant role in enhancing classification outcomes.

Table 1 Comparative Experiments on the ABIDE I Dataset

| Input Type | Model | ACC(%) | SEN(%) | SPEC(%) | AUC(%) |
|---|---|---|---|---|---|
| SWC | CNNLSTM | 66.56±2.74 | 72.06±4.50 | 58.57±3.13 | 70.65±4.00 |
| SWC | FBLSTM | 66.35±3.02 | 68.17±2.13 | 63.75±3.41 | 68.12±1.84 |
| SWC | STAGIN | 66.54±2.94 | 71.39±2.34 | 63.92±3.06 | 71.48±3.92 |
| SWC | BrainNetFormer | 68.86±3.12 | 72.49±3.46 | **66.12±2.65** | 74.02±3.92 |
| SWC+DEC | FE-STGNN | 65.90±3.02 | 69.19±1.53 | 64.48±3.12 | 71.57±2.92 |
| SWC+PS | CNNLSTM | 67.11±2.11 | 70.13±2.72 | 65.18±2.18 | 72.37±2.32 |
| SWC+PS | FBLSTM | 67.28±3.97 | 69.05±3.08 | 64.82±4.36 | 69.03±2.79 |
| SWC+PS | STAGIN | 66.89±2.98 | 70.81±2.41 | 64.33±3.18 | 70.87±3.79 |
| SWC+PS | BrainNetFormer | 66.73±2.39 | 70.20±2.96 | 65.02±1.40 | 70.29±2.21 |
| SWC+PS | SDCNN | 68.62±2.65 | 71.12±2.01 | 65.28±2.96 | 72.36±3.24 |
| SWC+PS | MAGCN | 66.30±2.93 | 70.39±3.50 | 64.34±4.38 | 72.09±4.27 |
| SWC+PS | FE-STGNN | 64.80±4.82 | 68.29±1.20 | 63.94±2.32 | 70.57±2.77 |
| SWC+PS | MSFL (Ours) | **70.98±2.52** | **74.61±4.61** | 65.58±5.32 | **76.24±1.62** |

Table 2 Comparative Experiments on the Rest-Meta-MDD Dataset

| Input Type | Model | ACC(%) | SEN(%) | SPEC(%) | AUC(%) |
|---|---|---|---|---|---|
| SWC | CNNLSTM | 63.70±3.54 | 63.03±7.21 | **65.31±3.85** | 66.12±7.62 |
| SWC | FBLSTM | 64.53±5.79 | 64.02±5.55 | 64.75±4.37 | 65.92±5.04 |
| SWC | STAGIN | 63.98±3.85 | 68.42±2.87 | 59.45±7.04 | 68.76±3.87 |
| SWC | BrainNetFormer | 64.20±2.66 | 64.02±4.95 | 63.49±2.74 | 69.51±3.03 |
| SWC+DEC | FE-STGNN | 66.09±1.63 | 69.23±1.01 | 64.30±1.93 | 72.89±0.90 |
| SWC+PS | CNNLSTM | 64.56±2.77 | 67.06±4.50 | 60.57±3.13 | 67.45±4.08 |
| SWC | FBLSTM | 65.53±4.29 | 66.02±5.31 | 64.13±4.27 | 66.87±5.40 |
| SWC | STAGIN | 64.08±4.10 | 69.24±5.27 | 60.15±4.21 | 69.90±4.71 |
| SWC | BrainNetFormer | 63.01±2.10 | 62.42±3.05 | 65.49±2.02 | 66.54±2.83 |
| SWC+PS | SDCNN | 66.05±3.65 | 67.65±2.35 | 64.22±3.56 | 72.65±2.96 |
| SWC+PS | FE-STGNN | 65.58±2.18 | 68.53±2.61 | 65.02±3.38 | 71.56±3.50 |
| SWC+PS | MAGCN | 60.32±3.29 | 59.48±5.23 | 63.19±4.32 | 62.92±6.16 |
| SWC+PS | MSFL (Ours) | **68.84±2.71** | **72.19±3.17** | 64.01±4.43 | **74.53±3.14** |

In the comparative experiments, the MSFL model demonstrated optimal performance across the three metrics of ACC, SEN, and AUC. This superior performance is likely attributable to the model's effective fusion of two feature extraction methods, SWC and PS, as well as the robust feature extraction capabilities of its backbone network. Additionally, it was observed that for single-channel models—particularly CNNLSTM, FBLSTM, and STAGIN—their dual-channel versions outperformed the original single-channel models on both datasets. This further confirms that incorporating PS features can significantly enhance classification results.

### 5.4 **Ablation Study of Phase Synchronization Preprocessing**

Prior to applying phase synchronization, the data must be preprocessed to smooth the signal and reduce excessive fluctuations. Common preprocessing methods include narrow-band filtering and Empirical Mode Decomposition (EMD). In this experiment, conducted using the REST-meta-MDD dataset, the model's feature fusion module was removed, and only the single-channel PS feature was used as input, while all other hyperparameters remained unchanged. EMD and narrow-band filtering were applied separately to preprocess the data. Specifically, the EMD method involved selecting and processing intrinsic mode functions from different frequency bands, whereas narrow-band filtering was applied across various frequency bands to evaluate its effect on classification metrics. The experimental results are presented in Table 3.

Table 3 Results with different Phase Synchronization Preprocessing

| Method | Frequency | ACC(%) | SEN(%) | SPEC(%) | AUC(%) |
|---|---|---|---|---|---|
| Not Pre-processed | None | 60.62±3.59 | 59.40±3.02 | 60.19±4.27 | 62.44±4.09 |
| EMD | 0.02Hz | 54.10±5.32 | 62.01±6.62 | 39.92±5.01 | 55.24±5.86 |
| EMD | 0.05Hz | 55.30±4.29 | 63.92±4.86 | 40.96±5.01 | 57.61±5.46 |
| EMD | 0.10Hz | 51.95±6.23 | 59.26±7.81 | 43.58±6.35 | 52.66±6.32 |
| EMD | 0.20Hz | 49.10±6.95 | 50.32±6.30 | 40.55±8.25 | 50.78±7.34 |
| Narrowband Filtering | 0.03~ 0.07Hz | 60.90±2.58 | 63.58±3.04 | 61.17±2.17 | 64.62±3.20 |
| Narrowband Filtering | 0.01~ 0.1Hz | 61.27±3.54 | 64.61±3.76 | 60.36±2.69 | 65.21±3.05 |
| Narrowband Filtering | 0.1~ 0.13Hz | **62.18±2.50** | **65.52±1.92** | **60.74±2.76** | **67.67±2.83** |

Although the EMD method is applicable in PS analysis, it may cause greater information loss in deep learning tasks, leading to poorer classification performance. In contrast, narrow-band filtering outperformed the absence of preprocessing. The best results were obtained using narrow-band filtering within the 0.01-0.13Hz frequency band. Therefore, for all subsequent experiments in this paper, phase synchronization preprocessing was performed using narrow-band filtering at 0.01-0.13Hz.

### 5.5 Ablation Study on the Feature Fusion Methods

To validate the effectiveness of the CDA module, this experiment compared its performance against single-channel inputs (using only SWC or PS features) as well as the effects of other attention mechanisms. The results are shown in Table 4, where "SWC" and "PS" represent single-channel input methods, "Concat" refers to the simple concatenation of the two features, and "CDA" represents a feature fusion module constructed using a single CDA module.

Table 4 Results on different Feature Fusion Methods

| Method | Acc (%) | Sen (%) | Spec (%) | AUC (%) |
|---|---|---|---|---|
| SWC | 66.44±3.42 | 69.54±4.23 | 63.42±4.64 | 70.12±3.85 |
| PS | 62.18±2.50 | 65.52±1.92 | 60.74±2.76 | 67.67±2.83 |
| CDA | 66.82±3.45 | 70.05±2.98 | 63.29±3.75 | 70.42±3.66 |
| Concat | 60.62±3.55 | 59.40±3.63 | 60.19±2.94 | 62.44±3.84 |
| Cross-Attention | 67.12±2.93 | 70.41±3.24 | **64.92±2.88** | 72.24±3.09 |
| Attention-Free | 65.65±2.35 | 68.22±3.56 | 62.96±1.95 | 69.61±2.56 |
| LAFF | 63.91±3.48 | 65.26±3.03 | 62.58±3.67 | 67.02±3.92 |
| HAT | 66.23±2.04 | 69.02±2.73 | 62.94±1.88 | 70.78±2.39 |
| ATFuse | 67.20±3.49 | 70.58±3.63 | 63.79±2.99 | 72.62±3.64 |
| MMViT | 67.75±2.85 | 70.32±3.09 | 64.12±2.73 | 73.10±3.04 |
| CA+CDA | **68.84±2.71** | **72.19±3.17** | 64.01±4.43 | **74.53±3.14** |

With the exception of the Concat and LAFF [33] methods, all other attention-based fusion

methods—including CDA, Cross-Attention, Attention-Free, HAT [34], ATFuse [35], and MMViT [36]—outperformed the single-channel inputs (SWC or PS). This indicates that fusing SWC and PS features provides a significant advantage. Among these, the feature fusion method proposed in this paper, composed of the CDA and CA modules, surpassed all other methods across the three metrics of ACC, SEN, and AUC.

### 5.6 Model Explanation and Feature Importance Analysis

To further investigate the mechanism by which feature importance affects model performance, we used SHAP to evaluate the importance of two types of features: sliding-window correlation and phase synchronization. The specific experimental procedure is as follows:

(i) The SHAP method was used to calculate the importance scores for each feature within the SWC and PS feature sets, respectively.

(ii) A grouped masking experiment was conducted based on the ranked scores. Group 1: Masking the top-x most important features in the SWC feature set; Group 2: Masking the top-x most important features in the PS feature set; Group 3: Jointly mask the top-x most important features from both the SWC and PS feature sets.

(iii) The trend in the model's accuracy (ACC) was quantitatively evaluated under different masking conditions.

The experimental results are presented in Figure 5, where the horizontal axis represents the number of masked features, and the vertical axis represents the model's accuracy. As the number of masked features increases from 10 to 200, the model's performance steadily declines, indicating that both SWC and PS features significantly contribute to classification accuracy. Notably, masking only the SWC features causes a faster decline in performance compared to masking only the PS features. Furthermore, jointly masking important features from both types results in the most substantial performance drop. For example, when 200 features were masked in all three groups, the accuracy of the joint masking group was 4% lower than that of the single-feature masking groups. This confirms that the two feature types provide complementary representations.

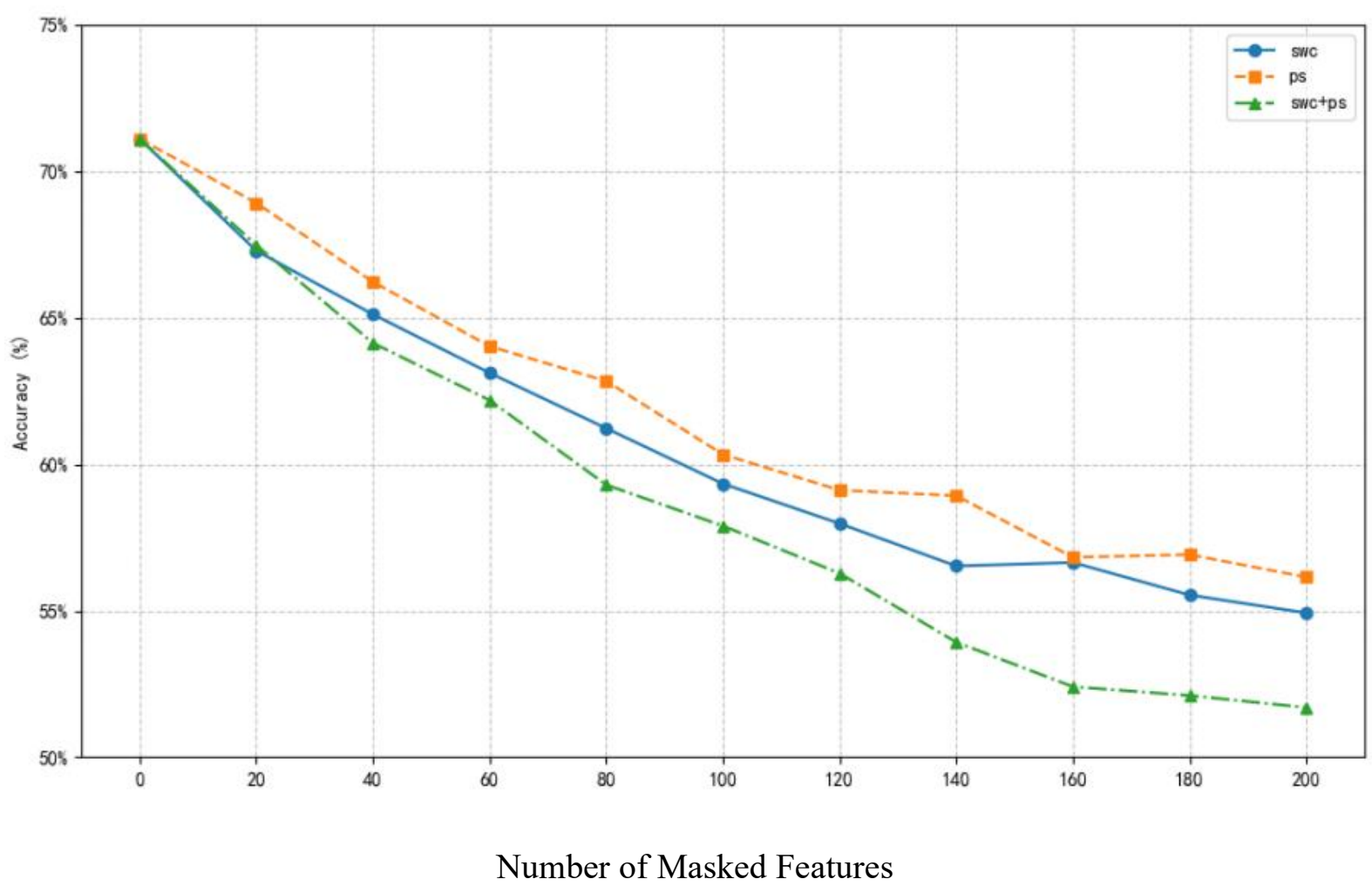

Fig. 5 Analysis of the Impact of Feature

Observing that the model's accuracy approached 50% when 180 connections were masked in the third group, this experiment compiled statistics on the top 180 feature connections ranked by SHAP contribution to further analyze the distribution patterns of SWC and PS features in brain functional connectivity. The analysis also examined the brain regions involved in these connections and identified functional connections that recurred across different feature types.

The results showed that, among the SWC features, brain regions 67 and 51 in the AAL brain atlas appeared most frequently, each occurring five times. In contrast, within the PS features, brain region 67 appeared three times, while brain regions 21 and 22 appeared six times each. Furthermore, among the 180 feature connections with the highest contributions, two connections—(39, 59) and (86, 87)—were present in both the SWC and PS features. This suggests that, although SWC and PS represent different measures of dynamic functional connectivity, they share some overlap in key connections and may jointly influence specific functional networks. Combined with the results of the masking experiment, these findings indicate that the complementary representation of these features is crucial to the classification model's decision-making process.

## 6 Conclusion

This paper proposes a deep network model, MSFL, which integrates two distinct types of dynamic functional connectivity features. The model's feature fusion module employs a cross-attention mechanism to extract complementary information from sliding-window correlation and phase synchronization features. Experimental results demonstrate that MSFL's classification performance on autism and depression datasets surpasses that of various existing baseline methods. Furthermore, ablation studies systematically evaluate the optimal preprocessing method for phase synchronization features and confirm the effectiveness of the feature fusion module. Finally, based on a feature importance analysis, the key brain regions involved in the top 180 functional connections, as well as recurring functional connections, are identified. These findings contribute to a deeper understanding of disease-related pathological mechanisms by combining the amplitude and phase of fMRI signals through dynamic functional connectivity.